# The Dependence of Alloy Composition of InGaAs Inserts in GaAs Nanopillars on Selective-Area Pattern Geometry

*Joshua N. Shapiro[1], Adam C. Scofield, Andrew Lin, Nicholas Benzoni,*

*Giacomo Mariani, and D. L. Huffaker*

UCLA Electrical Engineering and California Nano-Systems Institute

**Abstract:** GaAs nanopillars with 150 nm – 200 nm long axial InGaAs inserts are grown by MOCVD via catalyst-free selective-area-epitaxy (SAE). The alloy composition of the InGaAs region, as determined by room-temperature photoluminescence (PL), depends critically on the pitch and diameter of the selective-area pattern geometry. The PL emission varies based on pattern geometry from 1.0 µm to 1.25 µm corresponding to a In to Ga ratio from 0.15 to >0.3. This In enrichment is explained by a pattern dependent change in the incorporation rate for In and Ga. Capture coefficients for Ga and In adatoms are calculated for each pattern pitch. As the pitch decreases, these data reveal a contest between a synergetic effect (related to nanopillar density) that increases the growth rate and a competition for available material that limits the growth rate. Gallium is more susceptible to both of these effects, causing the observed changes in alloy composition.

---

[1] Corresponding author: jns@ee.ucla.edu

Catalyst-free selective-area-epitaxy (SAE) is a powerful technique for growing uniform arrays of nanopillars, and accurately controlling their location, diameter, and height[1-3]. The precise positioning and excellent uniformity of SAE nanopillars enables sophisticated device design with simple fabrication[4-7]. Despite the accuracy and control of nanopillar position and diameter, there are still important and unsolved problems related to how the SAE pattern geometry effects both the growth rate and the alloy composition of ternary materials such as InGaAs. These problems become particularly relevant when designing complex SAE patterns that attempt to leverage the periodic array of nanopillars for photonic crystals or plasmonic enhancement. Understanding how to utilize the SAE pattern to control both the growth rate and the material composition will improve the ability to engineer sophisticated and robust nanopillar opto-electronics.

There is a wide body of research that describes in detail how the vertical growth rate depends on the geometric properties of individual nanowires, but the study of collective effects in arrays of nanopillars grown by the catalyst-free method is only beginning[8-18]. Borgstrom et. al. investigated the effects of nanowire density for Au-catalyzed SAE and identified three overlapping growth regimes, the independent pillar regime, the synergetic regime, and the materials competition regime[9]. While their explanation is specific to catalyzed VLS epitaxy, the same trends appear in this work using catalyst-free SAE indicating a broader underlying principle. In this work, we investigate how the SAE pattern effects the growth rate of GaAs nanopillars, and how the pattern alters the relative incorporation rate of In and Ga atoms. The SAE pattern therefore enables controlled tuning of the alloy composition of axial InGaAs inserts embedded in GaAs nanopillars.

In this report GaAs nanopillars with axial InGaAs inserts and InGaP shells are grown by selective-area MOCVD. A shift and broadening in PL is observed as the selective-area pattern geometry varies. The observed change is attributed to the composition of the axial InGaAs segment. Compositional changes are possible if the selective-area pattern geometry alters the incorporation rate of Ga compared to In atoms. This relative incorporation rate is strongly correlated with the SAE pattern geometry. The incorporation rate for Ga is computed from the measured volume of the nanopillars as a pitch dependent capture coefficient, $\gamma_{Ga}$. The estimated InGaAs composition is then used to determine the capture coefficient, $\gamma_{In}$, for In. Finally, we discuss possible differences between In and Ga that can lead to the observed difference in their capture and incorporation.

The nanopillars are grown on semi-insulating GaAs 111B substrates with a 25 nm evaporated SiO$_2$ mask using techniques described previously[6,19-21]. The substrate is patterned with a 5x5 grid of nano-hole arrays by e-beam lithography and reactive-ion etching shown in Fig. 1a. Each pattern in the grid is a 50 $\mu$m x 50 $\mu$m array of nano-holes on a triangular lattice with a single diameter and pitch. There are a total of 25 patterns covering combinations of the diameters [40, 60, 80, 100, 120] nm and the pitches [0.2, 0.4, 0.6, 0.8, 1.0] $\mu$m. Each pattern in the grid is separated from its nearest neighbors by 100 $\mu$m. The nanopillar structure consists of a GaAs nanopillar with a single axial InGaAs insert and an axial InGaP shell. A schematic of the nanopillar structure is shown in Fig. 1b. The GaAs and InGaAs sections are grown in a hydrogen atmosphere at 730°C in a vertical flow MOCVD system at 60 Torr. The temperature is reduced to 600°C for growth of a 5-10 nm InGaP shell that efficiently passivates the structure[20,22].

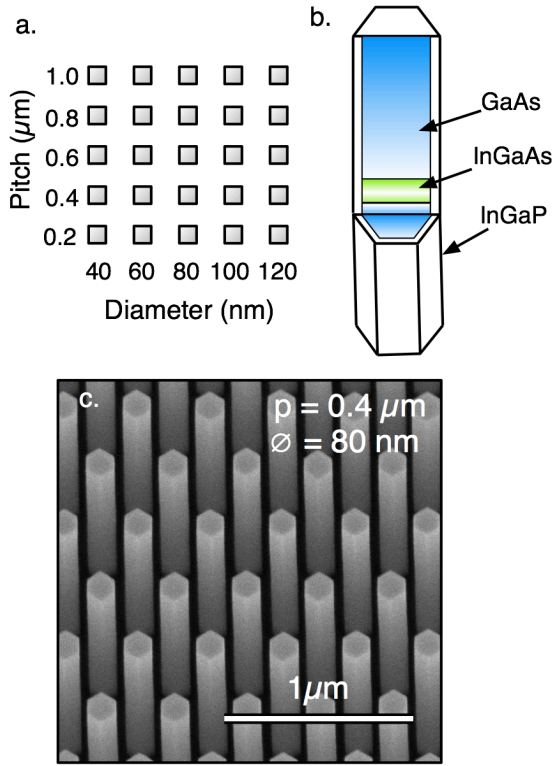

**Figure 1.** (a) Schematic of pattern layout. Each square is a 50 µm x 50 µm array of the specified pitch and diameter. (b) Schematic of nanopillar heterostructure with cutaway to show the InGaAs axial insert and InGaP shell. (c) Representative SEM of nanopillars.

Each of the 25 patterns are characterized by SEM and $\mu$-PL. A representative SEM image is shown in Fig. 1c. The final heights of the nanopillar range from 0.8 µm to 1.8 µm with variations from 5% to 15% depending on pattern. The nanopillar diameters are larger than the patterned hole diameters due to overgrowth, and variations in diameter are typically 3% for a given pattern. The SEM images are used to calculate the average nanopillar volume per unit area on each pattern, discussed in detail below.

Each axial InGaAs segment is 150-200 nm in length (see supporting information).

The normalized $\mu$-PL spectra from each of the SAE patterns are grouped by pitch in Fig. 2. All spectra show a peak at 916 nm (1.354 eV) with a tail extending to ~980 nm. This feature is commonly referred to as the 1.36 eV band, and is attributed to a point defect such as a Ga vacancy or a Cu impurity[23-26]. The emission from InGaAs is observed at wavelengths between 1000 nm and 1300 nm with a distinct variation from pattern to pattern. At a pitch of 0.2 $\mu$m the spectra show a red-shift with increasing diameter. When the pitch increases to 0.4 $\mu$m and 0.6 $\mu$m the spectra blue-shift, and maintain a mild diameter dependence. When the pitch increases further to 0.8 $\mu$m and 1.0 $\mu$m the PL emission becomes considerably weaker and more variable, but on average appears to red-shift again.

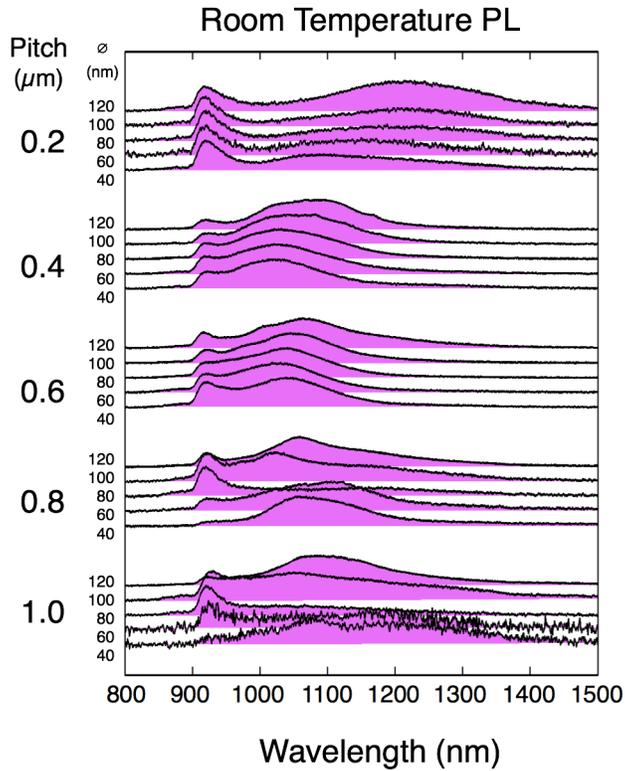

**Figure 2.** Normalized room temperature $\mu$-PL of each nanopillar array.

Neither quantum confinement nor strain are believed to have a significant effect on the PL emission. The inhomogenous broadening of the spectra is from variation in material composition, both within individual pillars and across the ensemble[19,27]. The length of the InGaAs inserts (> 150 nm) eliminates quantum confinement as a potential cause for the blue shift of the InGaAs peak. Strain can potentially blue-shift the emission, but should effect all pillars of similar diameters equally. These data show that diameter can either blue-shift or red-shift the spectra for a given pitch, so strain is an improbable cause of the blue-shift. The only remaining explanation for the shifting PL is changing material composition from pattern to pattern.

The PL peak wavelength and corresponding In composition of each spectra is plotted as a function of pitch and diameter in Fig 3a. The error bars represent the FWHM of each spectra. There is a 'U'-shape trend across the plot, with the largest and smallest pitch having the highest In composition ($X_{In}$). When the pitch is 0.2 $\mu$m, the In fraction reaches its highest values of $X_{In} > 0.30$. For intermediate pitch of 0.8 $\mu$m, 0.6 $\mu$m, and 0.4 $\mu$m, $X_{In}$ reaches its lower bound of ~0.15, but still exhibits variation with diameter. At the 1.0 $\mu$m pitch, the $X_{In}$ increases slightly again to between 0.2 and 0.3 depending on diameter. Compositional analysis of a few pillars agrees with the compositions estimated from PL (see supporting information). These data demonstrate a distinct correlation between $X_{In}$ and the SAE pattern geometry.

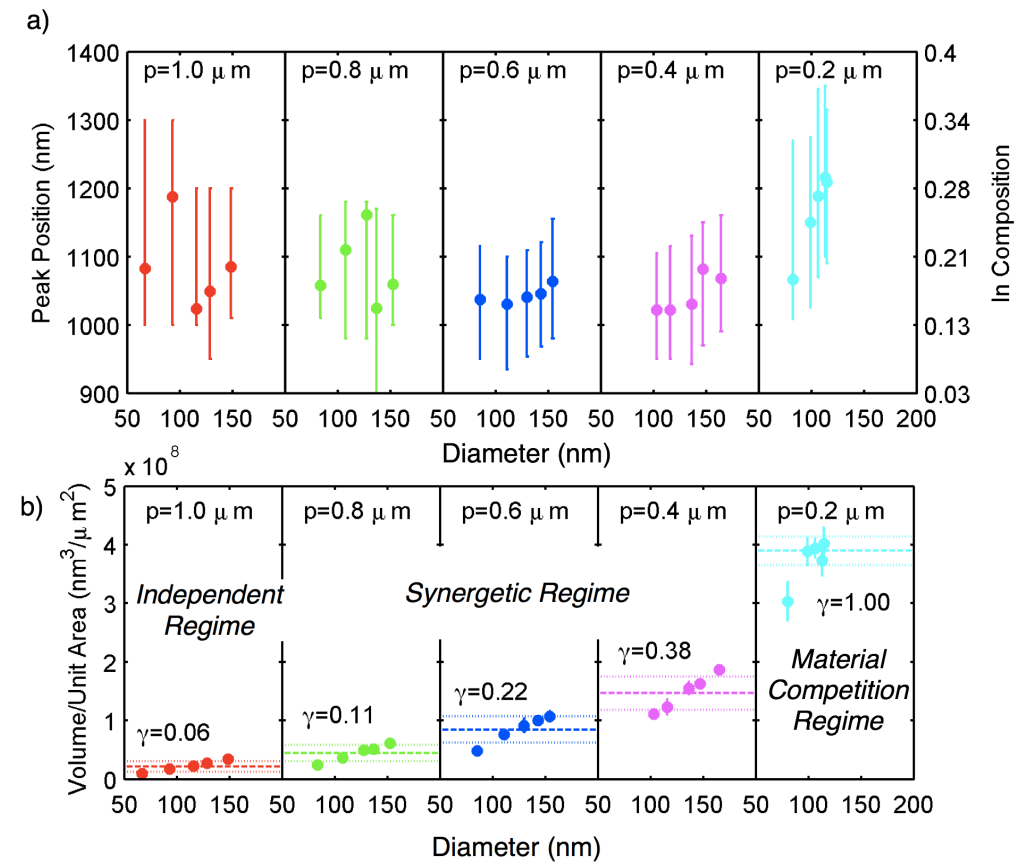

**Figure 3.** (a) The position of the PL peak and FWHM of the peak plotted as error bars vs. pitch and diameter. The corresponding InGaAs composition is plotted on the right axis. (b) The total volume per unit area of deposited material plotted versus pitch and diameter. Estimated capture coefficient is shown for each pitch as a dashed line, dotted lines indicate one standard deviation.

In order for the InGaAs composition to change with pattern geometry, the incorporation rate of Ga relative to In must change. In SAE, one cannot assume that the incorporation rate of incident atoms will be independent of the pattern. The diffusion length on the selective-area mask and the amount of exposed semi-conductor must play a crucial role in determining the capture probability for adatoms. Atoms adsorb on the growth mask and diffuse, but they do not necessarily encounter a nanopillar and incorporate. Intuitively, one can imagine that a sparse array of nanopillars captures a small fraction of material, while a dense array captures more material. In the limit where the nanopillar density approaches that of a thin-film, the capture probability should approach that of thin-film epitaxy, which for mass-transfer limited growth is equal to a sticking-coefficient of 1.

In this work, we determine this pattern dependent capture coefficienty by measuring the volume of material collected by each pattern. Furthermore, distinct capture coefficients for Ga and In atoms can be determined by comparing the volume measurement and the composition measurements. The pattern and species dependent capture coefficient measured here explains how the SAE pattern can be utilized to tune the InGaAs alloy composition.

The total nanopillar volume per unit area [$nm^3/\mu m^2$] is plotted versus diameter and pitch in Fig 3b. These volume data, V, for each SAE pattern are computed from the equation $V = hA/(p/2)^2$, where $h$ and $A$ are the height and cross-sectional area of 5 to 10 individual nanopillars measured by SEM, and $p$ is the pitch. These volume data reveal a strong relationship between pitch, diameter, and the total growth volume per unit area. This volume, V, increases steadily as the diameter increases and pitch decreases until the pitch equals 0.2 $\mu$m, where V plateaus at the mass-transfer limit. The volume is related to the rate at which atoms incorporate into the nanopillar array according to the equation $V = \chi t \Omega$, where $\chi$ is a (time-averaged) pattern dependent incorporation rate, $\Omega$ is the volume of the primitive unit cell, and $t$ is the growth time. This incorporation rate, $\chi$, is pattern dependent and proportional to the constant flux of atoms $\beta$ via a dimensionless capture coefficient, $\gamma$, that depends on the selective area pattern pitch and diameter. This leads to the following expression for volume in terms of $\gamma$,

$$V = \gamma \beta t \Omega. \qquad (1)$$

The capture coefficient for Ga, $\gamma_{Ga}$, is computed from Fig. 3b and Eq. 1 by setting $\gamma_{Ga} = 1$ where the volume plateaus at 0.2 $\mu$m pitch. This method ignores the minor contribution of In atoms to the total volume, and assumes $\beta$, $t$, and $\Omega$ are constant. We also do not attempt to determine the diameter dependence of $\gamma_{Ga}$, although there clearly is one, but instead limit our calculations to establishing the larger effect of pitch by averaging over all diameters for a given pitch. At large pitch, $\gamma_{Ga}$ is small and the nanopillars collect a small fraction of the available material. As the pitch decreases and the diameters increase, the nanopillars become more efficient at capturing atoms from the vapor until the smallest pitch, when $\gamma_{Ga}$ reaches its maximum value and saturates.

This trend of increasing incorporation rate as pitch decreases is the key evidence of synergetic growth in catalyst free nanopillars. The three growth regimes identified in Ref. [9], the *independent*, *synergetic*, and *materials competition regimes*, re-appear in these data. At large pitch, the growth occurs in the independent regime, where a single nanopillar collects some fraction of material within a critical radius that is small compared to the distance to a neighboring pillar. At small pitch, in the materials competition regime, the dense array of nanopillars collects all the available material from the vapor. The synergetic regime occurs at intermediate pitch, where neighboring nanopillars work together to collect an increasing fraction of the total material. The exact mechanism by which nanopillars cooperate to collect more material is beyond the scope of this work, but we hypothesize that when the pitch approaches the mean-free-path of an atom in the vapor, an atom can desorb from one pillar and adsorb on a neighbor without re-entering the vapor, effectively enhancing the collection efficiency.

Together the two plots in Fig. 3 present the relationship of both InGaAs alloy composition and incorporation rate to the SAE pitch and diameter. There is slight In enrichment at large pitch in the independent growth regime, but also generally broader PL spectra and less uniformity. A more substantial In enrichment occurs at small pitch, in the materials competition regime. For In enrichment to occur in the materials competition regime Ga must be subject to material competition while In is not. To further illustrate this difference between In and Ga, the capture coefficient $\gamma_{In}$ is computed.

The capture coefficient for In, $\gamma_{In}$, is calculated by equating the measured alloy composition with the ratio of the effective incorporation rates of In and Ga according to the equation,

$$X_{In} = \frac{\chi_{In}}{\chi_{In} + \chi_{Ga}} = \frac{\beta_{In}\gamma_{In}}{\beta_{In}\gamma_{In} + \beta_{Ga}\gamma_{Ga}}. \qquad (2)$$

Given the In:Ga ratio in the vapor phase, $X_{In}^{Vapor} = \beta_{In}/(\beta_{In} + \beta_{Ga}) = 0.48$ (calibrated from planar growth of $In_{0.53}Ga_{0.47}As$ lattice matched to InP at 600°C), the capture coefficient for In is computed from equations 1 and 2.

$$\gamma_{In} = \frac{X_{In}}{(1-X_{In})} \frac{(1-X_{In}^{Vapor})}{X_{In}^{Vapor}} \gamma_{Ga}. \qquad (3)$$

Fig 4. shows the calculated capture coefficients $\gamma_{In}$ and $\gamma_{Ga}$ as a function of pitch. The inset shows the mean In composition, $X_{In}$, for each pitch used to compute $\gamma_{In}$. Studying the capture coefficients for Ga and In, two effects emerge that appear to contribute to In enrichment. First, the onset of synergetic growth begins at a larger pitch for Ga compared to In. This is evident by examining the slopes of the two capture coefficients. The relatively mild slope of $\gamma_{In}$ between 1.0 and 0.5 $\mu$m compared to $\gamma_{Ga}$ shows that Ga atoms are more efficiently captured compared to In as the pitch decreases. Second, Ga enters the material competition regime before In. At 0.2 $\mu$m pitch, $\gamma_{Ga} = 1$ while $\gamma_{In}$ is still increasing sharply. Ultimately this demonstrates that the SAE pattern interacts with different species of atoms in measurably different ways that can be utilized to engineer and tune the final nanopillar structure.

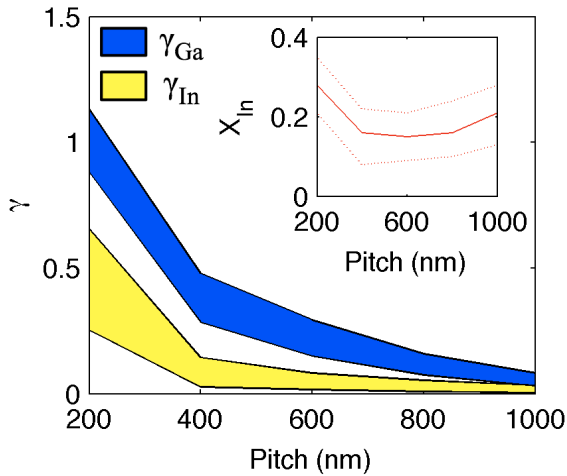

**Figure 4.** Calculated capture coefficients for gallium and indium from the volume measurements and the indium composition. *Inset)* The average indium composition vs. pitch, estimated from $\mu$PL.

Indium enrichment has been predicted and observed in nanowire epitaxy and selective area epitaxy of InGaAs in other work[27-31]. SAE of patterned microstrips have revealed indium enrichment, that is attributed to both a larger diffusion coefficient of In compared to Ga and a higher capture probability for Ga[28,29]. A study of the effect of nanowire density on InGaAs composition also attributed the enrichment to a longer diffusion length of In, albeit the substrate was bare GaAs (111), not a selective area mask[30]. Finally, InGaAs nanopillars grown by SAE show a modification of alloy composition with pitch that they hypothesize is caused by the growth rate being limited by the presence of Ga atoms[31].

A longer diffusion coefficient for In offers only a partial explanation of the observed phenomena, and actually may play only a minor role in the observed data. If the nanopillar arrays collect In atoms from a greater distance than Ga, then In enrichment can be expected when Ga becomes depleted first in the material competition regime. However the longer diffusion coefficient of In cannot explain the reduced In content in the synergetic regime.

Even though In has a longer diffusion length in the vapor and on the mask, the incorporation of this abundant In can be prevented by a high growth temperature (730°C). At this temperature, a significant fraction of In adatoms will rapidly desorb from the surface due to lower binding energies and generally lower cohesive energy of the InGaAs alloy[32]. So In may be plentiful in the vapor, but its tendency to be captured is quite low.

A simple geometric construction can explain the change of composition in the three growth regimes. Consider a circle surrounding each nanopillar as described in Ref. [18], and only atoms within the circle's radius are collected. At large pitch, these circles do not overlap leading to the independent growth regime. At smaller pitches, when the circles do overlap, atoms in the vicinity of overlapping circles are more likely to be captured, and are then available to any nanopillar in the neighborhood. Every nanopillar then collects atoms from the area defined by its own circle plus the overlapping portion of neighboring circles. Synergetic growth will result when the circles overlap, but the total number of atoms collected does not exceed the total number available. If the radius of a Ga collection circle is larger than for In, then the Ga collection circles will overlap at larger pitch than the In collection circles. The result will be that Ga will experience both synergetic growth and materials limited growth at larger pitch than In. Additional In available due to longer diffusion lengths, will further act to delay the onset of materials limited growth for In. While this simple geometric argument is not rigorous, it does qualitatively explain the observed growth regimes, and the observed change in alloy composition.

In conclusion, GaAs nanopillars with 150-200 nm axial $In_xGa_{1-x}As$ segments grown by SAE show a strong dependence of In composition on the SAE pattern geometry. The composition becomes In rich at large and small pitch, with a minimum In content at a pitch of 0.6 $\mu$m. This enrichment is explained by a pattern dependent capture coefficient that differs for In and Ga. The Ga capture coefficient increases more than In at intermediate pitch due to the onset of synergetic effects, and saturates at $\gamma_{Ga}$ = 1.0 when the pitch reaches 0.2 $\mu$m. In contrast, the capture coefficient for In does not exhibit synergetic effects until smaller pitches, and does not saturate for this experiment. A simple geometric argument can justify the observed data. This work demonstrates clear methods for using selective-area pattern geometry to control the alloy composition in nanopillar hetero-epitaxy. These techniques have potential in future opto-electronic devices that seek to engineer multiple distinct material compositions in a single growth.

This work is supported by NSF (through DMR-1007051, and DGE-0903720). The authors gratefully acknowledge the support from facilities and staff at the California Nano-Systems Institute at UCLA.

**Supporting Information:**

To acquire the spectra, the sample is mounted at the focus of a 50x near-infrared corrected microscope objective, and a $100\mu W$ HeNe laser is used to illuminate a spot ~$5\mu m$ in diameter. The PL is collected by the same microscope objective and detected with a cooled InGaAs focal plane array mounted on a 0.5 m focal length spectrometer.

Transmission electron microscopy and elemental analysis by energy dispersive x-ray spectroscopy (EDS) of selected nanopillars was performed to verify the structure. Figure S1a. shows dark field scanning TEM of a nanopillar from the pattern with 600 nm pitch and 180 nm diameter. The InGaAs insert (bright region) is 340 nm from the base of the pillar and approximately 160 nm long. Figure S1b. shows elemental mapping of the Ga-L peak from two nanopillars. The nanopillar from a pattern with 600 nm pitch is 20% In, 80% Ga in the InGaAs segment. The nanopillar from a pattern with 200 nm pitch is 30% In, 70% Ga in the InGaAs segment. These values agree with the estimated composition from PL.

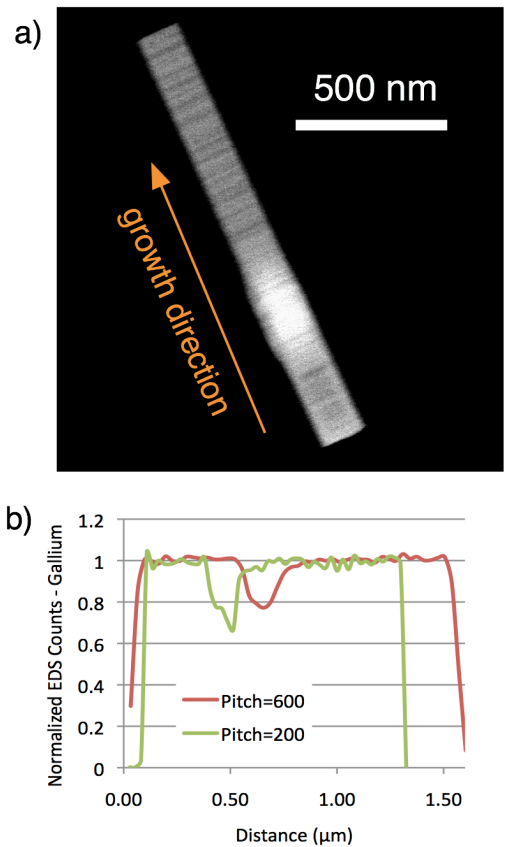

Figure S1. a) Dark Field STEM of a nanopillar from the pattern with 600 nm pitch. The InGaAs region is brighter than GaAs regions. b) Normalized EDS of the Ga-L peak for two nanopillars from different patterns.